# Casualty Dynamics in Wars and Terrorism and the Scale-Free Organization of Social Systems

**Ingo Piepers**

June 20th 2006

*Amsterdam, The Netherlands*
E-mail: *ingopiepers@gmail.com*

## Abstract

In this paper I propose a 'mechanism' for the explanation of power-law characteristics of casualty dynamics in inter-state wars, intra-state wars and terrorist attacks: the scale-free physical organization of social systems. Other explanations - self-organized criticality (Cederman, 2003) and the redistribution of total attack capabilities (Johnson et al. 2006) - do not provide a consistent framework for the power-law characteristics of casualty dynamics. The development in time of the power-law characteristics of casualty dynamics during wars and conflicts provides clues for the 'functioning' of social systems which are targeted, and/or for the (in)effectiveness and strategies of actors using force (violence) against these social systems.

**1. Introduction.**

In 1948 Richardson revealed that the number and size (in terms of casualties) of conflicts scales as a power-law; this is the case for *inter-* as well as for *intra-*state wars. Cederman proposes a self-organized critical (SOC) dynamic to explain Richardson's law (Cederman, 2003).
More recently Johnson et al. report that the power-law patterns of violence arising in three high-profile ongoing wars, and in global terrorism show remarkable universality. It is important to note that Johnson's et al. analysis looks at the pattern of casualties arising *within* a given war. Johnson's et al. theory treats the insurgent force as a generic, self-organizing system which is dynamically evolving through the continual coalescence and fragmentation of its constituent groups.
In my own research (Piepers, 2006) I have identified a power-law pattern in the size - size in terms of the number of Great Powers participating in Great Power wars in the period 1495-1945 - of wars. I argue that this power-law pattern is the result of a SOC-mechanism, resulting - through a punctuated equilibrium dynamic - in the development of the international systems towards a condition of increased stability, during the relative stable periods in between consecutive punctuations.
I assume that Richardson's and Johnson's et al. power-laws are related to the same 'basic' dynamic and mechanisms, however at different levels of analysis: Richardson's and Johnson's et al. power-laws reflect the fractal nature of ('the same') casualty dynamic. Furthermore, I assume that the power-law identified by Piepers is the outcome of a fundamentally different dynamic and mechanism (size in this power-law not defined in terms of casualties).

| **Comparison of theories** ||||
|---|---|---|---|
| | **Level of analysis** | **'Quantity'** | **Proposed mechanism** |
| Johnson's et al. | Within a war | Number of casualties | Continual coalescence and fragmentation |
| Richardson and Cederman | Inter- and intra-state wars | Number of casualties | SOC (Cederman) |
| Piepers | International system | Number of states | SOC |

*Table 1. Comparison of theories.*

Notwithstanding that some efforts have been made by Cederman (Cederman, 2003) and Johnson et al. (Johnson et al. 2006) to explain the typical power-law patterns in casualty dynamics, this typical statistical relationship has so far not been explained satisfactorily.

In this paper I argue that the explanations of Cederman and Johnson et al. are not consistent, and that these typical statistical relationships in the 'casualty dynamics' of social systems are closely related to the physical - scale-free - organization of social systems. In other words: the power-laws which characterize the casualty dynamics of various conflicts 'reflect' the basic organization of these social systems. This explanation implies that force (violence) - at (conflict) system level - is to a high degree, applied at random.

I will address the following subjects and issues: (1) the explanations of Cederman and Johnson et al. for respectively Richardson's law and power-law patterns in three ongoing wars and global terrorist attacks, (2) an alternative framework based on the assumption that the typical power-law patterns in casualty dynamics result from the scale-free physical organization of social systems, (3) various arguments taking the edge off Cederman's explanation and off the assumptions and model Johnson et al. have constructed, and (4) some implications of the 'scale-free theory' discussed in this paper.

**2. Self-organized criticality and the redistribution of total attack capabilities.**

Cederman and Johnson et al. respectively argue that the power-law patterns in the casualty dynamics of conflicts and wars can be attributed to a SOC-mechanism and the constant redistribution of capabilities of attack units. In this section I will describe the most important arguments and logic of both theories.

The main logic in Cederman's explanation of *inter*-state wars - discussed in *"Modeling the Size of Wars: From Billiard Balls to Sandpiles"* (Cederman, 2003) - pertains to how geopolitical instability changes strategic calculations; Cederman refers to this mechanism as 'context activation' (Cederman, 2003, 138). Furthermore Cederman argues - referring to Gilpin (Gilpin, 1981) - "that change tends to be driven by innovation in terms of technology and infrastructure". According to Cederman "such cases of technological change may facilitate both resource extraction and power projection" (Cederman, 2003, 137). Cederman assumes that technological change and contextual activation characterize the dynamics of inter-state wars and that these factors are responsible for Richardson's law (Cederman, 2003, 144). According to Cederman does this 'framework' constitute a self-organized critical system, with technological change as an important driving force of the international system.

Johnson et al. report in their research paper *"Universal patterns underlying ongoing wars and terrorism"* (Johnson et al, 2006.): ".... a remarkable universality in the patterns of violence arising in three high-profile ongoing wars[1], and in global terrorism. Our results suggest that these quite different conflict arenas currently feature a common type of enemy, i.e. the various insurgent forces are beginning to operate in a similar way regardless of their underlying ideologies, motivations and the terrain in which they operate" (Johnson et al. 2006, 1)

Johnson et al. provide in this paper a microscopic theory to explain their main observations. This theory treats the insurgent force as a generic, self-organizing system which is dynamically evolving through the continual coalescence and fragmentation of its constituent groups (Johnson et al. 2006, 1). Furthermore Johnson et al. "suggest that the dynamical evolution of these various examples of modern conflict has less to do with geography, ideology, ethnicity or religion and much more to do with the day-to-day mechanics of human insurgency; the respective insurgent forces are effectively becoming identical in terms of how they operate" (Johnson et al. 2006, 2). And their microscopic mathematical model "represents the insurgent force as an evolving population of attack units whose destructive potential varies over time. Not only is the model's power-law behavior in excellent agreement with the data from Iraq, Colombia and non-G7 terrorism, it is also consistent with data obtained from the recent war in Afghanistan. These findings suggest that modern insurgent wars tend to be driven by the same underlying mechanism: the continual coalescence and fragmentation of attack units" (Johnson et al. 2006, 2). Johnson et al. describe this mechanism in more detail as follows: "Our model proposes that the insurgent force operates as a dynamically evolving population of fairly self-contained units, which we call 'attack units'. Each attack unit has a particular 'attack strength' characterizing the number of casualties which they typically arise in an event involving this attack unit. As time evolves, these attack units either join forces with other attack units (i.e. coalescence) or break up (i.e. fragmentation). Eventually this on-going process of coalescence and fragmentation reaches a dynamical steady-state which is solvable analytically, yielding a power-law with coefficient α = 2.5. The combination of these empirical and analytical findings suggest that similar distributions of attack units are emerging in Colombia, Iraq and in non-G7 global terrorism, with each attack unit in an ongoing state of coalescence and fragmentation" (Johnson et al. 2006, 6).

Johnson et al. argue that "these empirical and analytical findings lead us to speculate that power-law patterns will emerge within any modern asymmetric war which is being fought by loosely-organized insurgent groups" (Johnson et al. 2006, 7).

### 3. An alternative mechanism: the 'underlying' scale-free organization of social systems.

Contrary to Cederman and Johnson et al., I assume that the more or less random use of force (violence) 'against' or within a social system *with a scale-free organization*, rationally - inevitably - results in a casualty dynamic with power-law characteristics. This means that a deviation from a 'perfect' casualty power-law can most probably be attributed to (1) 'deviations' in the scale-free organization of the targeted social system and/or (2) the (in)effectiveness of the actor(s) applying force against the targeted social system[2].
Evidence is available for the scale-free (physical) organization of social systems: city populations follow a power-law pattern (Newman, 2005), as well as transport systems, and the Internet is a (virtual) scale-free network (Barabási et al. 2003).
In order to establish if 'organizations' which are involved in or related to conflicts have scale-free physical organizations, I have studied in more detail the alliance structure (network) of the international system, and the typical structure of military organizations.

*a. The scale-free structure of formal alliances.*

Based on the Correlates of War dataset I have analyzed the 'structure' of the network of alliances wit Great Power (n = 196) involvement in the international system in the period from 1816 until 2000 (Gibler et al. 2004). Two observations are neglected (two alliances with an exceptional size of respectively 20 en 22 participants, resulting in a 'heavy tail' of the distribution): it is evident that the organization of the international system has scale-free characteristics.

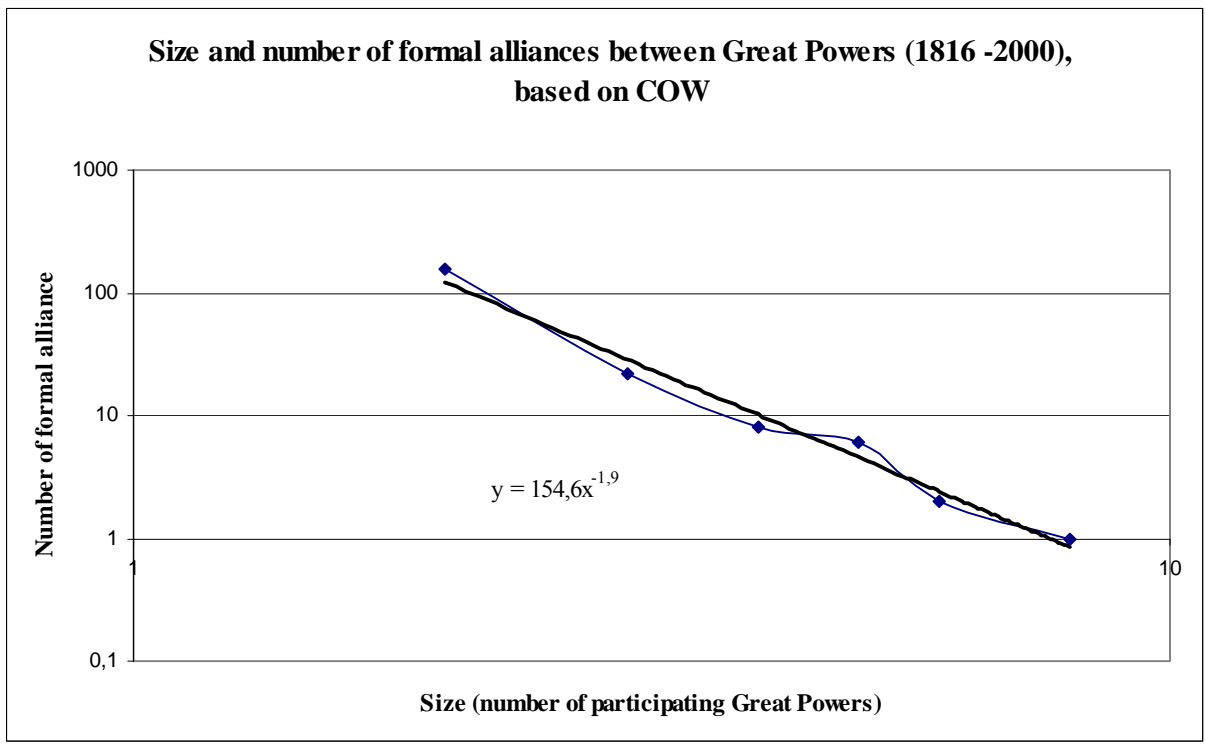

*Figure 1. Scale- free characteristics of the organization of the international system (source: Correlates of War dataset, n = 196)).*

I do *not* consider the existence of this power-law to be sufficient proof for the universality of scale-free (physical) organization structures of social systems, furthermore it is important to note that the alliance organization is related to the organization of the international system and not to the level of 'conflict systems'. However I do consider these scale-free characteristics of the international system yet another indication that social systems tend to organize as scale-free structures.

*b. The fractal organization structure of military organizations.*

Military organizations have a fractal - scale-free - structure. This organizational structure is to a high degree 'self-similar'. Obviously is this scale-free structure of military organizations the most effective structure, ensuring an optimal trade off between (fighting) effectiveness and vulnerability. In this example I assume that a military organization - irrespective of its 'level' of hierarchy - consists of three manoeuvring - fighting - units, a support unit, a logistical unit and a command function.

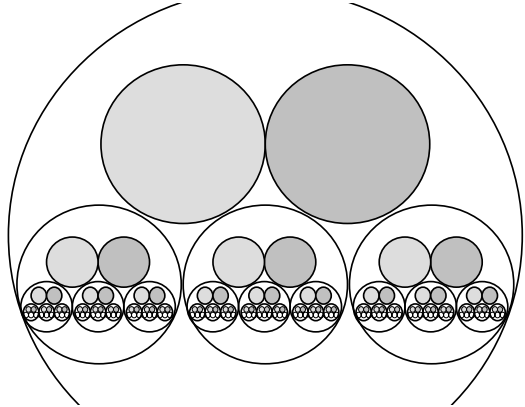

*Figure 2. The fractal - self-similar - structure of military organizations.*

In order to determine if it is a plausible assumption that the casualty dynamics (within a war) reflect the structure of military organizations, I have (roughly) 'compared' the coalition casualty dynamics in Iraq (from March 19th 2003 until May 21st 2005), with the typical structure of military organizations (with five hierarchical levels).
The casualty figures are based on the "Iraq Coalition Casualty Count" (www.icasulaties.org), and do only include the casualties of Coalition Forces[3]. In this 'model' the theoretical military unit consists of three subunits with a fighting task (manoeuvre units)[4]. The line corresponding with the distribution of the number and size of fighting units only gives an indication of the relative distribution of (sub) units at various hierarchical levels.
It is interesting to note that it seems a plausible assumption that the (military) casualty dynamics reflect the structure of military organizations; however further research is required.

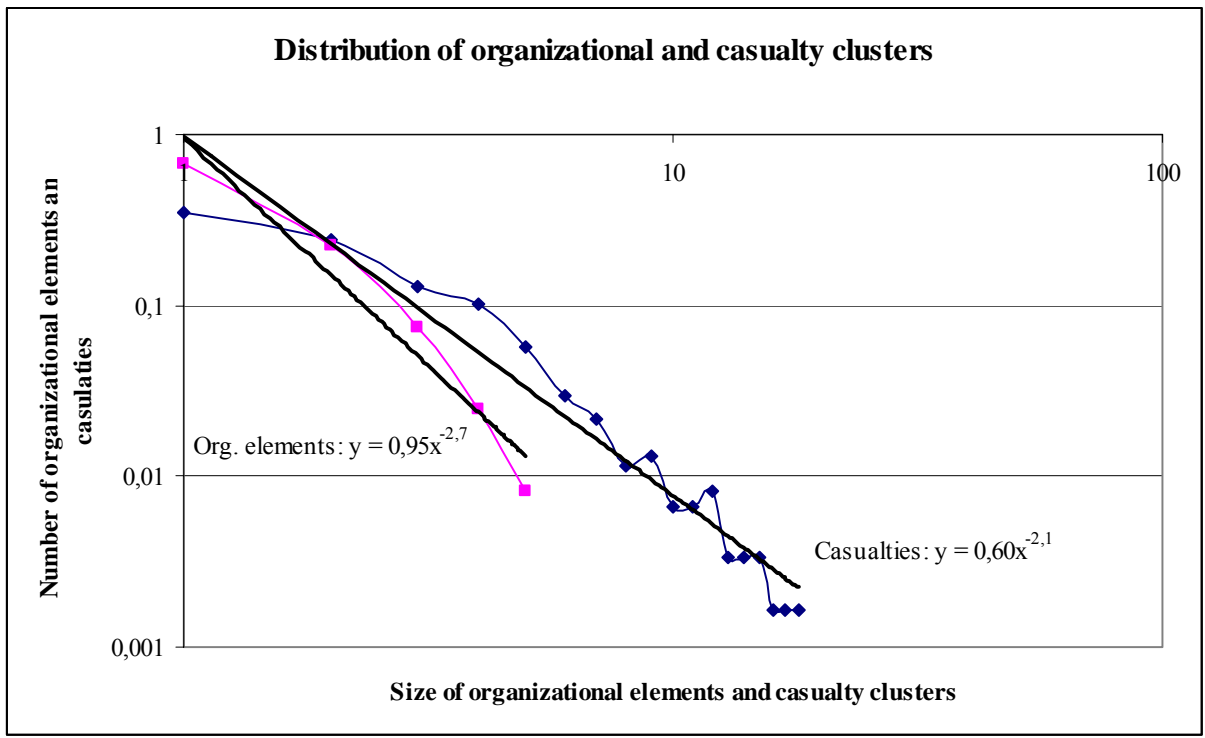

*Figure 3. The distribution of organizational and casualty clusters.*

**4. A critical evaluation of Cederman's theory.**

In this section I discuss Cederman's theory and I argue that his framework is based on inconsistent assumptions, and that in Cederman's model two levels of analysis are confused: casualty dynamics at the level of conflict systems, and inter-state dynamics at the level of the international system.

Cederman's explanation is not consistent for the following reasons.

- Cederman restricts his article and explanation to *inter*-state wars, neglecting the existence of a more or less similar power-law for the severity and size of *intra*-state wars. The role of the state, context activation and technological change are important elements of Cederman's theory, but do not apply to *intra*-state wars. Cederman's theory can not - for that reason - provide an explanation for the power-law patterns in the casualty dynamics of *intra*-state wars. This issue is not addresses by Cederman.
- Cederman does not identify the 'critical point' in 'his' SOC-system; this is an essential 'component' of such a system: a SOC-system often oscillates 'around' a critical point.
- Typically, in a SOC system a driving force results in a gradual - more or less regular - build up of tension, enabling a separation of timescales (Sornette, 2004, 404). Technological change is not a gradual process and can for that reason not qualify as the 'driving force' of a SOC-system[5]. I do not consider the simulations with the agent based model - constructed by Cederman on the basis of his assumptions - proof for the existence of a mechanism resulting in casualty dynamics with power-law characteristics.

**5. A critical review of Johnson's et al. assumptions and model.**

In this critical evaluation of Johnson's et al. assumptions and model I will focus on (1) the data used and the interpretation of this data, (2) assumptions underlying Johnson's et al. model, and (3) Johnson's et al. conclusions.

*a. Inconsistencies in data and data-interpretation.* Johnson et al. use various databases (Johnson et al. 2006, 36). These databases include guerilla, civilian and military casualties (Johnson, 2006, 36-40). I will now elaborate on two inconsistencies in the interpretation of Johnson et al. related to the data used in their research.

- In some data killings of (Afghan) civilians *attributed to coalition forces* are included (Johnson et al. 2006, 40) and obviously (?) do not result in a significant distortion of the power-law patterns Johnson et al. have identified. However, coalition forces apply force according to fundamentally different principles and methods (tactics) then irregular forces. If a close relationship exists between the modus operandi of irregular forces and the power-law patterns in casualty dynamics - as Johnson et al. argue - then a significant distortion of the power-law should be unavoidable.
- Johnson et al. rightly observe that the Iraq war breaks into two phases (Johnson et al. 2006, 39): a conventional phase (March 20, 2003 to April 30, 2003) and an irregular/insurgent phase (from May 1st, 2003 onwards). Jonhson et al. find "that we get similar results if we use the combined data from both phases". In my opinion is this finding not consistent with Johnson's et al. conclusion that the typical power-law pattern with a coefficient of $\alpha \approx 2.5$ is closely related to the modus operandi of the irregular forces operating in Iraq (and other conflicts). If such a close relationship exits (between the typical modus operandi of irregular forces and the power-law pattern identified by Johnson et al.), then the data related to the conventional phase should result in a (significant) distortion.

It is important to note that these inconsistencies conflict with Johnson's et al. conclusions, but are consistent with my own framework, based on the assumption that the scale-free organization of social systems (military and non-military) rationally results in power-law patterns in casualty dynamics, irrespective of how and by whom force is applied.

*b. Assumptions conflict with reality.* In below table I discuss some of the assumptions Johnson et al. make in their microscopic theory (mathematical model): in their assumptions Johnson et al. neglect organizing and (non) conventional military principles.

| **A critical evaluation of assumptions** | | |
|---|---|---|
| | **Assumptions** | **Comment** |
| 1 | The attack strength of an attack unit represents the number of casualties which the attack unit would *typically* inflict in a conflict event[6] (Johnson et al. 2006, 7 and 14). The attack strength of an attack unit in the model is fixed. | The attack strength of an attack unit is not fixed (constant) and to a high degree context dependent, dependent on the moral of the unit, on available resources (weapons, explosives, information, etc.), and on the (constantly) evolving tactics of these units. |
| 2 | The on-going process of coalescence and fragmentation reaches a dynamical steady-state (Johnson et al. 2006, 6). | In most wars - starting as irregular wars - organizations of these irregular forces evolve towards more conventional structures. Such a development is necessary in order to be able to defeat (militarily) the (conventional) forces and avoid endemic conflicts. A steady state - as defined by Johnson et al. - is an exception of only short duration. |
| 3 | The total attack strength for the entire insurgent force would change fairly slowly over time (Johnson et al. 2006, 13). | Attack strength - the ability to inflict casualties - does not evolve fairly slowly over time and is to a high degree dependent on the tactics of attack units and various - complex - factors. Attack strength - especially of irregular forces - in asymmetric conflicts develops very 'freakish' and surprising. Irregular forces constantly adopt other tactics to exploit vulnerabilities; (ir)regular organizations are learning organizations. |
| 4 | The process of coalescence and fragmentation is a stochastic process. | This process is not (totally) stochastic; some simple 'guidelines' do apply: the strategy and objectives of the irregular forces and principles regarding (operational) security and the massing of attack strength. Probably a chaotic attractor influences this dynamic as well (Piepers, 2006). |
| 5 | The individuals constituting an attack unit are probably connected by location or connected by some communication system (Johnson et al. 2006, 13). | This definition raises the question if coordinated attacks (in various locations) should be considered a single attack. What effect does this 'correction' have on the power-law pattern? |
| 6 | Attack units with different attack strength will continually mutate via coalescence and fragmentation yielding a 'soup' of attack units with a range of attack strengths. Johnson et al suggest that such a process might also underpin the acts of terrorism in non-G7 countries and that such terrorism is characteristic of some longer-term 'global war' (Johnson, et al. 2006, 16) | Johnson et al. suggest with this explanation that their model applies to 'global terrorism' as well, in other words that this model provides an explanation for the power-law coefficient (2.5) for non-G7 countries terrorism. However, these non-G7 terrorist attack units do not coordinate their attacks on a global scale and a process of coalescence and fragmentation does not exist. These 'terrorist organizations' often 'just' consist of some individuals or independent 'cells' (probably 'inspired' by a common cause). |
| 7 | Johnson et al. suppose that each attack unit has a given probability of being involved in an event, regardless of its attack strength (Johnson et al. 2006, 16). | Johnson et al. suggest in their model a relationship between size and attack strength (Johnson et al. 18). However, the probability of an attack unit being involved in an event is size-dependent: this is due to the complexity and risks of large(r) scale attacks: the larger the size of an attack unit the smaller the probability that this unit will be involved in an event. |
| 8 | Attack units will fragment, coalesce or remain unchanged (Johnson et al. 2006, 18). In case of coalescence the attack strength of the newly formed attack unit is the sum of the attack strengths of its parts. | Johnson et al. neglect in their calculations positive and negative synergy effects of coalescence and fragmentation. Coalescence should result in positive synergy, 'adding' to the attack strength of the attack unit(s) involved. |

| | **A critical evaluation of assumptions** <br> **(Continued)** | |
|---|---|---|
| | **Assumptions** | **Comment** |
| 9 | The probability of choosing an attack unit for coalescence is proportional to its attack strength. According to Johnson et al. it becomes more risky - and less worthwhile - to combine attack units as the attack units get smaller (Johnson et al. 2006, 18 and 28). | Smaller attack units will coalesce more easy and faster, without endangering (operational) security. For this reason smaller attack units will coalesce (and fragment) more often. |
| 10 | Size of attack units and attack strength are positively related. | This is - typically - not the case in asymmetric warfare (as far as the insurgents are concerned). The most deadly attacks are often conducted by individuals or by very small groups (car bombs and road side bombs). |

*Table 2. A critical evaluation of assumptions underlying Johnson's et al. model.*

*c. Conclusions, model and data are not consistent.* I do not doubt that Johnson's et al. model - microscopic theory - does 'generate' power-laws, with a certain range for α. However, I do doubt if this model realistically reflects the 'workings' of conflict systems (this is a similar shortcoming as in Cederman's agent-based model). At least one theory - Johnson's et al. and/or Cederman's - (in my opinion both) is incorrect; after all both theories try to explain the same phenomenon, but use fundamentally different concepts.

Jonhson et al. note that the theoretical results are consistent with, and to some extent explain, the various power-law exponents found for conventional wars, terrorism in G7 countries, and terrorism in non-G7 countries (Johnson et al. 2006, 32). Some (additional) comments:

- *Conventional wars*. There is and was - as suggested by Johnson et al. (referring to the power-law exponent of 1.8 for conventional wars) *no* "tendency toward building larger, robust attack units with a fixed attack strengths *as in a conventional army*, as opposed to attack units with rapidly fluctuating attack strengths as a result of frequent fragmentation and coalescence processes". Conventional military organizations evolve(d) towards more flexible structures and tactics (Smith, 2005, English, 1984).
- *Terrorism in G7 countries*. The power-law exponent 1.7 for G7 terrorism can as Johnson et al suggest be interpreted - on the basis of Johnson's et al. assumptions and model - as an even stronger tendency for robust units to form and an increased tendency to form larger units - or equivalently - to operate as part of a larger organization. However, this interpretation is based - as discussed - on the (incorrect) assumption that the attack strength and the size of attack units are closely related (in irregular warfare). Furthermore, it is not clear how this development is related to the assumption that larger units are more vulnerable to attack and fragment with a higher probability.

**6. Implications.**

In this paper I propose a mechanism for power-law patterns in casualty dynamics in inter-state wars, intra-state-wars and terrorism, based on the assumption that social systems - military and non-military - have scale-free physical organization structures. I assume that the casualty dynamics with power-law characteristics reflect this basic organizational structure of social systems. This line of thought has the following implications:

- At the level of a conflict system force (violence) is applied at random and that the presumed rationality of actors in the theories of Cederman and Johnson et al. does not have a significant impact on the casualty dynamics of conflict systems.
- Deviations from a power-law pattern in casualty dynamics can be attributed to deviations in the basic organization of the targeted social system or to the (in)effectiveness of attack units in the application of force;
- The evolution of the power-law exponent can be attributed to the reaction (in structural terms) of the social system to the targeting by military units and/or terrorist organizations, and/or to the in- or decreased effectiveness of organizations applying force against these social systems.
- The power-law characteristics of casualty dynamics are related to the basic (scale-free) organization of social systems - the 'receiving' side - in the conflict system and not to the dynamics of attack units - the supply side - applying force against social systems, as Johnson et al. suggest.

- Only one mechanism is responsible for the power-law characteristics of casualty dynamics in various types of war/conflict. (It is a doubtful line of thought of Johnson et al. to 'link' power-law patterns within wars exclusively with "modern asymmetric wars", because it can be argued that Richardson's law covering wars in the period from 1820-1945, in fact constitute a series of (large scale) battles *within a period of more or less constant war.* The fact that the casualty dynamics show power-law patterns at both levels of analysis is consistent with the fractal nature of this phenomenon. I argue that the statistical characteristics of these "modern asymmetric wars" are not an exemption, and in accordance with previous wars, despite the fact that these - or some - wars have a different power-law coefficient).

Various mechanisms have been suggested for the scale-free organization of networks; e.g. preferential attachment, a mechanism assuming growth of networks and the preference of individuals to connect with other well connected individuals (Albert, 2002). Another (complementary?) explanation for the scale-free organization of social networks could be related to the reduced vulnerability of this type of networks for random attacks (the random 'removal' of individuals), resulting in an evolutionary advantage and an overall greater survivability of the system.

Further research is necessary to test these three theories.

**End notes:**

[1] Iraq, Afghanistan and Colombia (Johnson et al. 2006, 2).

[2] Johnson et al. have taken the physical organization of social systems into account as an explanatory factor. However this hypothesis is tested against Colombia data and is ("resoundingly") rejected (Johnson et al. 2006, 6). I doubt if this hypothesis is tested rigorously at the right level of analysis of the Colombian society (the level at which individuals and groups are targeted).

[3] I have excluded two events with relatively high casualty rates, respectively 34 and 37 casualties. I consider these two events 'noise'.

[4] I assume that this organization consist of 3 brigades, 9 battalions, 27 companies, 81 platoons and 243 groups.

[5] In the paper "*Dynamics and Development of the International System: A Complexity Science Perspective*" (Piepers, 2006), I argue that the international system during the period 1495-1945 does have SOC-characteristics. However this is a fundamentally different SOC-system then defined by Cederman, at a different level of analysis. In my opinion Cederman has confused various levels of analysis: trying to explain dynamics at the level of conflict systems (casualty dynamics), with the typical dynamics - SOC-dynamics - between states constituting the international system.

[6] Johnson et al. define attack strength "$s_i$ of a given attack unit *i*, as the average number of people who are typically injured or killed as the result of an event involving attack unit *i*. In other words, a typical event (e.g. attack or clash) involving group *i* will lead to the injury or death of $s_i$ people" (Johnson et al. 2006, 14).